\documentclass[conference]{IEEEtran}

\usepackage{pifont}
\usepackage{amsmath}
\usepackage{varwidth}
\usepackage{multirow}
\usepackage{adjustbox}
\usepackage{enumerate}

\ifCLASSINFOpdf
\else
\fi

\begin{document}
\title{Requirements Model for Cyber-Physical System}	
	
%\author{\IEEEauthorblockN{Md. Masudur Rahman}
%	\IEEEauthorblockA{
%		Institute of Information Technology\\
%		University of Dhaka, Bangladesh\\
%		bit0413@iit.du.ac.bd
%	}
%	\and
%	\IEEEauthorblockN{Md. Rayhanur Rahman}
%	\IEEEauthorblockA{
%		Institute of Information Technology\\
%		University of Dhaka, Bangladesh\\
%		ray.scarlet@gmail.com
%	}
%}
\author{\IEEEauthorblockN{Md. Masudur Rahman\IEEEauthorrefmark{1} and
		Naushin Nower\IEEEauthorrefmark{2}}
	\IEEEauthorblockA{Institute of Information Technology, University of Dhaka, Dhaka, Bangladesh}
	\IEEEauthorblockA{\IEEEauthorrefmark{1}bit0413@iit.du.ac.bd, 
		\IEEEauthorrefmark{2}naushin@iit.du.ac.bd
	}}
% make the title area
\maketitle
	
\begin{abstract}
%\boldmath
The development of cyber-physical system (CPS) is a big challenge because of its complexity and its complex requirements. Especially in Requirements Engineering (RE), there exist many redundant and conflict requirements. Eliminating conflict requirements and merged redundant/common requirements lead a challenging task at the elicitation phase in the requirements engineering process for CPS. Collecting and optimizing requirements through appropriate process reduce both development time and cost as every functional requirement gets refined and optimized at very first stage (requirements elicitation phase) of the whole development process. Existing researches have focused on requirements those have already been collected. However, none of the researches have worked on how the requirements are collected and refined. This paper provides a requirements model for CPS that gives a direction about the requirements be gathered, refined and cluster in order to developing the CPS independently. The paper also shows a case study about the application of the proposed model to transport system.
\end{abstract}
\IEEEpeerreviewmaketitle
	
\section{Introduction}
% no \IEEEPARstart
%The needs for data sharing
Cyber-Physical Systems (CPS) integrate computing and communication capabilities with monitoring and control of entities in the physical world. These systems are usually composed by a set of networked agents, including - sensors, actuators, control processing units and communication devices. As a result, CPS conforms a complex system which leads difficulties in the requirements engineering process because requirements are also conflicting, redundant and and complex. The redundant and complex requirements increases the time and cost of developing the system. 
\\
Elicitation phase  plays a vital role of the  requirements engineering process because it will act as a blueprint of the whole developments life-cycle. Elicitation phase is conformed by the activities - collecting requirements from multiple stakeholders and categorizing those requirements based on priority into normal, expected and exciting requirements. Normal requirements are those which must be accomplished by the developers. This type of requirements should be explicitly agreement between stakeholders and development team. Expected requirements are such those are necessary for the system to work correctly. This requirements are also explicitly documented. The third part, exciting requirements are extra features those can be provided by developers to give extra satisfaction to users/clients. This requirements are known as WOW factor in software requirement engineering.
\\
The requirements for cyber-physical system are complex to identify and refinement of  those requirements is also a difficult task as CPS is a complex and distributed system. The more appropriate way the requirements get collected and refined, the easier the development of the system. Existing works focus on requirements after collected, but they neither focus on the collection process nor the refinement process of the requirements for CPS.
\subsection*{Contribution}
We propose a generic requirements engineering model for cyber-physical systems that serves as the basis for requirements engineering at elicitation phase and documentation having suitable categories based on priorities. This model helps developers to work independently in each of the subsequent development phases - designing, implementation, testing, etc. to develop CPS system as the requirements model forms multiple clusters based on CPS computation, communication, control, etc.
\subsection*{Paper Organization}
The paper is organized as follows. Section II discusses the existing work related to requirements for CPS, Section III introduces the proposed requirements model, while Section IV discusses a case study based on the model. Finally, Section V concludes the paper with a outlook of future work.

\section{Related Work}
Physical components of CPS are qualitatively different from object-oriented software components. As a result, requirements engineering process for CPS also varies with respect to traditional software development process. Some existing researches have been worked on the requirements engineering process or only on some specific requirements that should be consider during CPS development. The existing works can be divided into two major categories: Time related requirements and generic requirements
\subsection{Time Related Requirements}
Time is an important requirement for developing CPS system which should be make the system deterministic and predictable \cite{lee2008cyber}. This paper has examined the challenges in designing CPS systems and in particular has raised the question of whether today’s computing and networking technologies provide an adequate foundation for CPS. Edward A Lee has advised in the paper to incrementally improve technologies and computer OS (Operating System) architecture to ensure time requirement.
\\
\\
CPSs need to conduct the transmission and management of multi-modal data generated by different sensor devices. In \cite{kang2008real}, a novel information centric approach for timely, secure realtime data services in CPSs has been proposed. In order to obtain the crucial data for optimal environment abstraction, L. H. Kong et al. \cite{kong2010optimizing} have studied the spatio-temporal distribution of CPS nodes. H. Ahmadi et al. \cite{ahmadi2010congestion} has presented an innovative congestion control mechanism for accurate estimation of spatio-temporal phenomena in wireless sensor networks performing monitoring applications. A dissertation on CPSs discusses the design, implementation, and evaluation of systems and algorithms that enable predictable and scalable real-time data services for CPS
applications \cite{kang2009adaptive}.
\\
However, time is not exactly predictable due to the hardware configuration and programming language nature. This only requirement can not ensure and focus the whole requirements engineering process.
\subsection{Generic Requirements}
Requirements engineering for systems of systems faces extremely distributed requirements engineering activities that involve a multitude of stakeholders. This often results in isolated requirements engineering approaches which, in turn, lead to requirements that can hardly be integrated with the other units of the SoS in order to keep them consistent \cite{penzenstadler2012requirements}. Penzenstadler et al. have provided a content model for CPS system requirements engineering. 
\\
M{\`e}ndez et al. \cite{fernandez2013improving}  has shown fundamentals and lessons learned in requirements engineering from developing a meta
model for artefact-orientation. They have reported on a case
study with a street traffic management business unit from
Siemens on the application of an artefact-based requirements
engineering approach \cite{mendez2011case}.
\\
Berenbach et al. \cite{geisbergertum} has described
requirements engineering (RE) artifact modeling and name the key components
to be a measurable reference model, a process tailoring
approach, and respective process guidelines. Berenbach describes each of these elements and suggests practices for their elaboration.
In \cite{ringert2014requirements}, Jan Oliver et al. described design requirements for robotics systems only using I/O automata.
\\
However, none of these researches have not focused on how the requirements will be collected and how the requirements are organized as well as optimized in the requirements engineering process for cyber-physical systems.
\section{Proposed Requirements Model}
Requirements engineering for CPS is a challenging task as there are many requirements conflicting each other of complex CPS systems. Elicitation part of the requirements specification is focused in this paper as it is one of the most vital parts of the engineering process. Collecting requirements and removing the conflicts depends on how appropriately the requirements are gathered and categorized. In addition, it will make development easier by reducing time and cost at elicitation phase as well as it will affect positively in the subsequent development phases. The following requirements model is developed in this paper for CPS system requirements engineering process at elicitation phase (figure 1).
\begin{enumerate}
	\item Business Domain Analysis
	\item Collect System Requirements
	\item Remove Conflict Requirements and Grouping the Requirements as –
		\begin{enumerate}
			\item Normal/Essential 
			\item Expected
			\item Exciting
		\end{enumerate}
	\item Cluster the Requirements	
\end{enumerate}
% Figure
\begin{figure}[ht]
	\centering
	\includegraphics[width = 3.5in,height=3.70in,angle = 0] {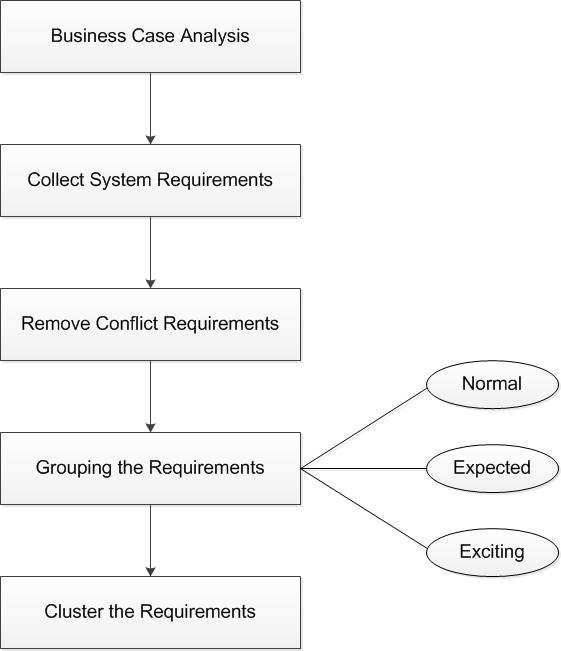}
	\caption{Flow Chart of Requirements Model for CPS}
	\label{RequirementModelFlowChart}
\end{figure}
Each part of the model is described briefly in the following contents.
% explain each item briefly
\subsection*{1. Business Case Analysis}
Business case (BC) of a system consists of certain contents such as - problem statement, mission/vision statement, Objectives of the system, rationality of the approach, benefits of all relevant stakeholders, performance measurement, risk management, work plan and time-line, team role, cost estimation and fund, etc. The topics of a BC provide us a useful source for analyzing requirements after collecting the requirements from stakeholders. Even we can generate such necessary requirements from mission/vision and objectives part of the BC that some of those might not  be collected during collaboration with stakeholders. 
\subsection*{2. Collect System Requirements}
In the inception phase during collaboration with stakeholders, system requirements to be collected and listing those requirements. The list contains conflicting requirements such as - response time vs output data volume, and common requirements such - system protection from other accesses and authentication. These types of requirements should be managed carefully in order to optimize the requirements so that it will be more easier to handle the requirements as well as it will reduce a lot of development time and cost.
\subsection*{3. Remove Conflict Requirements and Grouping the Requirements}
In the requirement list, common and conflict requirements are found. It is necessary to remove the conflict requirements and merge the common ones. As a result, the list would be so optimized that it will be easier to handle the requirements and fulfilling those ones for such a complex cyber-physical systems. After getting the optimized list, it is necessary to refine the requirements int order to grouping those ones into the following three categories.
 \begin{enumerate}[(a)]
 	\item Normal/Essential: 
 	This type of requirements is explicitly described by the stakeholders and must be met in the system development.
 	\item Expected: This type of requirements is collected from the system and business case analysis and through collaboration with stakeholders. Although these are not explicitly defined by stakeholders, these should be identified clearly and explicitly as without these requirements, the system will not functionally be correct or complete. So these requirements must also be met in the CPS product.
 	\item Exciting: This category of requirements are provided by the developers only as an excitement feature. This is not mandatory to work the CPS system correctly, but it will give extra satisfaction to clients. This is known as WOW factor in Software Requirements Specification. These requirements are completely dependent on the developments team whether to provide or not.
 \end{enumerate}
 The requirements can also be divided into two groups:
 \begin{enumerate}[(a)]
 	\item Functional: The three categories described above fall into the functional group as each of the requirements will be a function (or feature) of the system. 
 	\item Non-functional: It is also important to identify non-functional requirements at the elicitation phase for CPS. This requirements might be collected through business case analysis. Reliability, throughput, response time, performance, etc. are the example of non-functional requirements.
 \end{enumerate}
\subsection*{4. Cluster the Requirements}
Since CPS is a complex system having multiple modules like - computation, communication and control, the requirements are needed to be clustered based on the modules in order to distributing the workloads among the developers. As a result, each developer can work independently of other modules. So development goes faster. Three tire architecture can be applied here to cluster the requirements - 
\begin{enumerate}[(a)]
	\item Requirements related to business logic (computation logic)
	\item Requirements related to user interface (UI)
	\item Requirements related to data access/database
\end{enumerate}

\section{Case Study: Transport System}
We provide a case study of transport system to apply the proposed requirements model. To understand the model easily, we assume that the system consists of a set of cars and a traffic signal on the center in which several roads are connected. The scenario is pictorially represented in figure 2\footnote{Full resolution of picture online: \\
	$http://1.bp.blogspot.com/_CcCQJKOcXa0/Sv1onj6HKfI/AAAAAAAAAZM/OOmlh9k0Qo0/s400/Interactive-Vehicle-Communication-lg.jpg$}. The cars receive signals by their sensors from the traffic signal locating at a predetermined distance. These cars also sense information about the road's two sides range and other cars at a specific distance range.
\begin{figure}[ht]
	\centering
	\includegraphics[width = 3.5in,angle = 0] {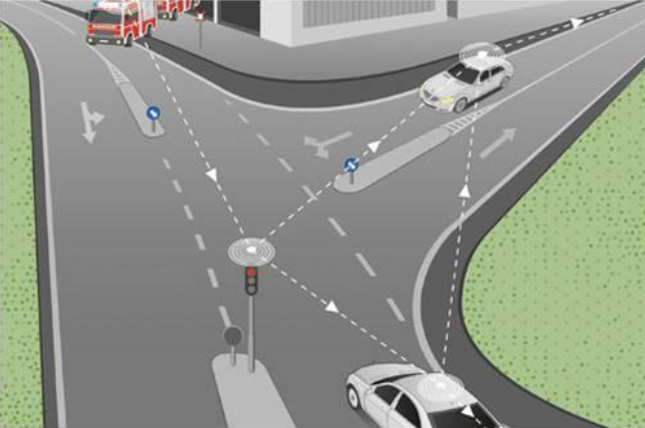}
	\caption{Transportation System}\label{fig:1b}
	\label{Transportation}
\end{figure}
According to the above business scenario, a portion of requirements collection and refinement using the proposed model have described in the following parts.
\subsection*{1. System Requirements} 
The following requirements are listed through collaborating with stakeholders.
\begin{enumerate}[(i)]
	\item Object identification range 10 meters
	\item Sense data through headlight and backlight of a car
	\item Avoid Crash
	\item Move the car when no object is found
	\item Stop the car when an object is found in front of the car or get signal to stop
	\item Wait when an object is at a specific distance
	\item Detect the road line/range
	\item Count time
	\item Accident should not occur
	\item Alert when other cars get closed to the car
	\item Alarm to find any object suddenly
	\item Control car speed
	\item Monitor vehicle speed
	\item Automatically track vehicle location in real-time
	\item Automatically determine schedules and efficient route 
	\item Provide safety and security 
	\item Move fast at free road
\end{enumerate}
\subsection*{2. Remove Conflict Requirements and Grouping the Requirements}
To refine the requirements stated above, we need to merge the common requirements and remove the conflict ones. The requirements (iii) and (ix) both are related to avoiding crash and hence these should be merged into one requirement. The requirements (xvi) and (xvii) are conflicting as it is not possible to move fast and get higher safety. So we emphasis on higher safety instead of fast moving. Requirements (x) and (xi) point to the same requirement alert when necessary. Requirements (xii) and (xiii) can be merges into one monitor and control speed. However, the final requirements are categorized into the three categories:
\subsubsection*{Normal Requirements}
\begin{enumerate}[(i)]
	\item Object identification range 10 meters
	\item Sense data through headlight and backlight of a car
	\item Avoid Crash
	\item Move the car when no object is found
	\item Stop the car when an object is found in front of the car or get signal to stop
	\item Wait when an object is found at a specific distance
	\item Detect the road line/range
	\item Provide safety and security 
\end{enumerate}
\subsubsection*{Expected Requirements}
\begin{enumerate}[(i)]
	\item Alert when other cars get closed to the car
	\item Count time
	\item Monitor and control car speed
	\item Automatically track vehicle location in real-time
	\item Automatically determine schedules and efficient route 
	\item Provide safety and security 
\end{enumerate}
\subsubsection*{Non-functional Requirements}
The following non-functional requirements are identified through business case analysis.
\begin{enumerate}[(i)]
	\item Reliability whether it work properly or not
	\item Safety measurement
	\item Performance measurement with respect to real time data collection
\end{enumerate}
\subsection*{3. Cluster the Requirements} 
Cluster the requirements based on sensors' data collection and functional business logic as so the developers can work independently to each cluster.
\subsubsection*{Cluster-I: Requirements Related to Data Collection}
\begin{enumerate}[(i)]
	\item Object identification range 10 meters
	\item Sense data through headlight and backlight of a car
	\item Detect the road line/range
	\item Monitor car speed
	\item Automatically track vehicle location in real-time
\end{enumerate}
\subsubsection*{Cluster-II: Requirements Related to Functional Business Logic}
\begin{enumerate}[(i)]
	\item Avoid Crash
	\item Move the car when no object is found
	\item Stop the car when an object is found in front of the car or get signal to stop
	\item Wait when an object is found at a specific distance
	\item Provide safety and security 
	\item Alert when other cars get closed to the car
	\item Count time
	\item Control car speed
	\item Automatically determine schedules and efficient route 
	\item Provide safety and security 
\end{enumerate}
\section{Conclusion and Future Work}
\subsection{Conclusion}
The requirements process for CPS is a challenging and important task for CPS on which the subsequent development phases depends. The proposed requirements model provides a strategy of collection and refinement of the requirements. Clustering requirements is a vital part of the model as it provides developers to work independently of other clusters of requirements. As a result, the model facilitates parallelism which reduces development time and cost dramatically.  
\subsection{Future Work}
The proposed model provides a complete approach of requirements engineering at elicitation phase. However, there is a possibility to work on scenario based modeling including usecases generation using this requirements model as future work. 

%\bibliographystyle{ieeetr}
%\addcontentsline{toc}{chapter}{Bibliography}
%\bibliography{mybib}

\begin{thebibliography}{9}
	\bibitem{lee2008cyber} 
	Lee, Edward A. "Cyber physical systems: Design challenges." Object Oriented Real-Time Distributed Computing (ISORC), 2008 11th IEEE International Symposium on. IEEE, 2008.	
	\bibitem{kang2008real} 
	Kang, Kyoung-Don, and Sang H. Son. "Real-time data services for cyber physical systems." Distributed Computing Systems Workshops, 2008. ICDCS'08. 28th International Conference on. IEEE, 2008.	
	\bibitem{kong2010optimizing}
	Kong, Linghe, Dawei Jiang, and Min-You Wu. "Optimizing the spatio-temporal distribution of cyber-physical systems for environment abstraction." Distributed Computing Systems (ICDCS), 2010 IEEE 30th International Conference on. IEEE, 2010.
	\bibitem{ahmadi2010congestion}
	Ahmadi, Hossein, Tarek F. Abdelzaher, and Indranil Gupta. "Congestion control for spatio-temporal data in cyber-physical systems." Proceedings of the 1st ACM/IEEE International Conference on Cyber-Physical Systems. ACM, 2010.
	\bibitem{kang2009adaptive}
	Kang, Woochul. Adaptive real-time data management for Cyber-Physical Systems. University of Virginia, 2009.
	\bibitem{penzenstadler2012requirements}
	Penzenstadler, Birgit, and Jonas Eckhardt. "A requirements engineering content model for cyber-physical systems." Requirements Engineering for Systems, Services and Systems-of-Systems (RES4), 2012 IEEE Second Workshop on. IEEE, 2012.
	\bibitem{fernandez2013improving}
	Fernández, Daniel Méndez, and Roel Wieringa. "Improving requirements engineering by artefact orientation." International Conference on Product Focused Software Process Improvement. Springer Berlin Heidelberg, 2013.
	\bibitem{mendez2011case}
	Fernández, Daniel Méndez, et al. "A case study on the application of an artefact-based requirements engineering approach." Evaluation \& Assessment in Software Engineering (EASE 2011), 15th Annual Conference on. IET, 2011.
	\bibitem{geisbergertum}
	Geisberger, Eva, et al. "Requirements engineering reference model (REM)." Technische Universität München, München (2006).
	\bibitem{ringert2014requirements}
	Ringert, Jan Oliver, Bernhard Rumpe, and Andreas Wortmann. "A Requirements Modeling Language for the Component Behavior of Cyber Physical Robotics Systems." arXiv preprint arXiv:1409.0394 (2014).
\end{thebibliography}
	
	% that's all folks
\end{document}